\documentclass[submission, Phys]{SciPost}

\usepackage{amssymb}
\usepackage{latexsym}

\usepackage{url}
\usepackage{xcolor}
\definecolor{newcolor}{rgb}{.8,.349,.1}

\usepackage{graphicx}	
\usepackage{amsmath}	
\usepackage{amssymb}	
\usepackage{bm} 
\usepackage{enumitem} 

\newcommand\bonsai{\textsc{Bonsai-SPH}}
\newcommand\bonsaig{\textsc{Bonsai}}
\newcommand\phantomx{\textsc{Phantom}}

\begin{document}


\begin{center}{\Large \textbf{
Bonsai-SPH: A GPU accelerated astrophysical Smoothed Particle Hydrodynamics code
}}\end{center}

\begin{center}
Jeroen  {B\'edorf}\textsuperscript{1,2},
{Simon {Portegies Zwart}}\textsuperscript{1},
\end{center}

\begin{center}
{\bf 1} Leiden Observatory, Leiden University, P.O. Box 9513, 2300 RA Leiden, the Netherlands
\\
{\bf 2} Minds.ai, Inc., Santa Cruz, the United States
\\
* bedorf@strw.leidenuniv.nl
\end{center}

\begin{center}
\today
\end{center}

\section*{Abstract}
{\bf
We present the smoothed-particle hydrodynamics simulation code,
\bonsai, which is a continuation of our previously developed
gravity-only hierarchical $N$-body code (called \bonsaig).
The code is optimized for Graphics Processing Unit (GPU)
accelerators which enables researchers to take advantage of
these powerful computational resources. \bonsai\ produces 
simulation results comparable with 
state-of-the-art, CPU based, codes, but using an order of 
magnitude less computation time.  The code is
freely available online and the details are described in this work.
}

\vspace{10pt}
\noindent\rule{\textwidth}{1pt}
\tableofcontents\thispagestyle{fancy}
\noindent\rule{\textwidth}{1pt}
\vspace{10pt}





\section{Introduction}

Smoothed Particle Hydrodynamics (SPH) is a particle based simulation
method for fluids. The method was introduced in astrophysics in 1977
by ~\cite{1977AJ.....82.1013L} and ~\cite{1977MNRAS.181..375G}, and
since then used in numerous simulations of astrophysical phenomena
addressing a wide range of phenomena including (but far from complete)
processes such as supernovae
feedback~\cite{doi:10.1111/j.1365-2966.2006.10785.x}, supernova shell
morphology~\cite{2016ComAC...3....2R}, galaxy
dynamics~\cite{doi:10.1111/j.1365-2966.2012.21733.x}, evolution of
disks around and accretion onto black
holes~\cite{2012A&A...545A.127R,2018MNRAS.477.4308R}, star
formation~\cite{doi:10.1111/j.1365-2966.2006.10495.x,2017ApJ...835..287G},
mass-transfer in binary and triple-star systems
\cite{2014MNRAS.438.1909D}, stellar
evolution~\cite{2010MNRAS.402..105G}, stellar collisions
\cite{1976ApL....17...87H,2016MNRAS.456.3401P}, and planet formation
\cite{2013MNRAS.429..895P,2018Icar..314..121S}.

There are several different SPH prescriptions, each with its own
specific advantages and disadvantages.  Many of the slightly different
implementations of generally the same algorithm are available in the
public scientific domain, or private.  Well known public
implementations include TreeSPH~\cite{1989ApJS...70..419H},
Phantom~\cite{phantom_paper}, and
Gadget~\cite{2001NewA....6...79S,2005MNRAS.364.1105S}. Here, we
present, yet another public SPH code. The main difference between
other codes and our implementation is that we use Graphical Processing
Units to accelerate the calculations.  This enables a faster execution
compared to existing codes and therefore the simulation of larger
models within the same wall-clock time-frame.  We deliberately do not
develop another SPH prescription, but take advantage of the large body
of previous work as described in recent reviews
~\cite{2010ARA&A..48..391S, PRICE2012759, doi:10.1146/annurev-fluid-120710-101220} and references therein.

SPH is one of several commonly used methods to simulate fluid dynamics, another 
well known group of method are the Eulerian (mesh/grid-based) solvers. Each 
method has its own advantages and disadvantages. For example, one SPH's key 
strengths is that it is self-adaptive, because it is based on particles. With 
mesh-based methods one has to resort to adaptive meshes in order to achieve a 
similar level of flexibility and computational efficiency. The result is a limited 
domain range for models 
that contain low density regions. A commonly cited disadvantage of SPH is
the method's difficulty with capturing shocks. This is because the 
method has inherent zero intrinsic dissipation and one therefore has to add 
dissipative terms such as artificial viscosity~\cite{2012ASPC..453..249P}. On 
the other hand shocks are handled naturally in grid-based methods~\cite{2007MNRAS.380..963A}. However, when the proper viscosity settings 
are chosen SPH is capable of handling shocks, such as those occurring in the 
Kelvin-Helmholtz instability test~\cite{phantom_paper}.

Recently there have been various methods introduced that combine the best parts 
of SPH and mesh based methods and combine this into a so called moving-mesh 
method. In this method the grid cells move with the fluid 
flow~\cite{10.1111/j.1365-2966.2009.15715.x, 2012MNRAS.425.3024V, 
2015MNRAS.450...53H,2019arXiv190904667W}. Although these methods are 
computationally intensive none of them are accelerated by GPUs although some 
work is being done on getting the underlying methods to run on GPU 
processors~\cite{doi:10.1145/3272127.3275092}.

Given the large amount of literature available that discuss the advantages and 
disadvantage of the various hydrodynamic methods we refer the reader to the 
following excellent reviews but point the interested reader to the references in 
this work, in particular to \cite{2010ARA&A..48..391S, PRICE2012759, 
2012ASPC..453..249P, doi:10.1146/annurev-astro-082214-122309, 
doi:10.1063/1.5068697} and do not go into further detail here unless it directly 
touches upon the goals of this paper, accelerating the SPH method with the use 
of Graphical Processing Units.

Most of the previous work related to hydrodynamics and GPUs has
focused on mesh-based codes. For example, in a version of {\tt
  FLASH}~\cite{0067-0049-131-1-273, Lukat201614} the authors
accelerate the (non-hydrodynamic) gravity computation using GPUs. The
optimization of the gravity module helps speeding up the code, but the
hydrodynamics modules are still running on the CPU and form a major
fraction of the total compute time. Another approach is taken in {\tt
  GAMER}~\cite{ 2010ApJS..186..457S,2017arXiv171207070S}, here the
authors implement the grid based hydro computations on the GPU and
present results that are qualitatively comparable to {\tt FLASH}, but
using over an order of magnitude less compute time.  Previous work
directly related to GPU accelerated astrophysical SPH methods is
limited, however there is previous work done on non-astrophysical SPH
methods. For example {\tt GPUSPH}~\cite{herault_sph_2010} and {\tt
  DualSPHysics}~\cite{CRESPO2015204}, which are CFD codes that use
the SPH algorithm.

This lack of available codes and the fact that in SPH the organization
of the particles can be done efficiently with the use of hierarchical
data-structures (trees) \cite{1989ApJS...70..419H,
  2001NewA....6...79S} has motivated us to develop a new optimized SPH
code.  We build this new SPH code on top of our existing gravity only
simulation code
\bonsaig~\cite{2012JCoPh.231.2825B,Bedorf:2014:PGT:2683593.2683600},
which uses an hierarchical data-structure to compute gravity and as
such can naturally be extended to simulate fluid dynamics.
\bonsai\ has been developed to take advantage of Graphics Processing
Units (GPUs) accelerators and all the actions related to the tree are
executed on the GPU.  This results in a high performance, scalable,
simulation code that enables us to perform the simulation of very
high-resolution models in reasonable time~\cite{2018MNRAS.477.1451F}.
With the increased availability of GPUs in the world's largest
supercomputers~\cite{top500} it will allow researchers to take
advantage of this new infrastructure and as such perform larger
simulations than possible before.

This work is organized as follows, in Sect.~\ref{sect:SPH} we shortly
introduce the SPH method and the specific version of SPH that we use
in this work, in Sect.~\ref{Sect:Implementation} we describe our
implementation, in Sect.~\ref{Sect:Results} we present the results and
in Sect.~\ref{Sect:Conclusion} we present our conclusion and
suggestions for future work.


\section{Smoothed Particle Hydrodynamics}
\label{sect:SPH}
\subsection{Overview}
In SPH the simulated fluid is discretised into a set of particles
where each particle has a position, $p$, velocity, $v$, and mass,
$m$. This enables SPH to solve the hydrodynamics equations in the
Lagrangian form. In contrast a grid code, as for example,
ENZO~\cite{2014ApJS..211...19B} solves the hydrodynamical equations
using a Eulerian form. Where the difference is that in the Lagrangian
form you change the properties of the individual particles as the
fluid moves. While in the Eulerian form you change the properties of
fixed locations as the fluid passes through these locations.  More
information about the difference between these two methods as used in
astrophysics can be found in~\cite{2007MNRAS.380..963A} and references
therein.  See also~\cite{zwart2018astrophysical} for a comparison
between the various methods.

\begin{figure}
   \begin{center}
    \includegraphics[width=0.8\columnwidth]{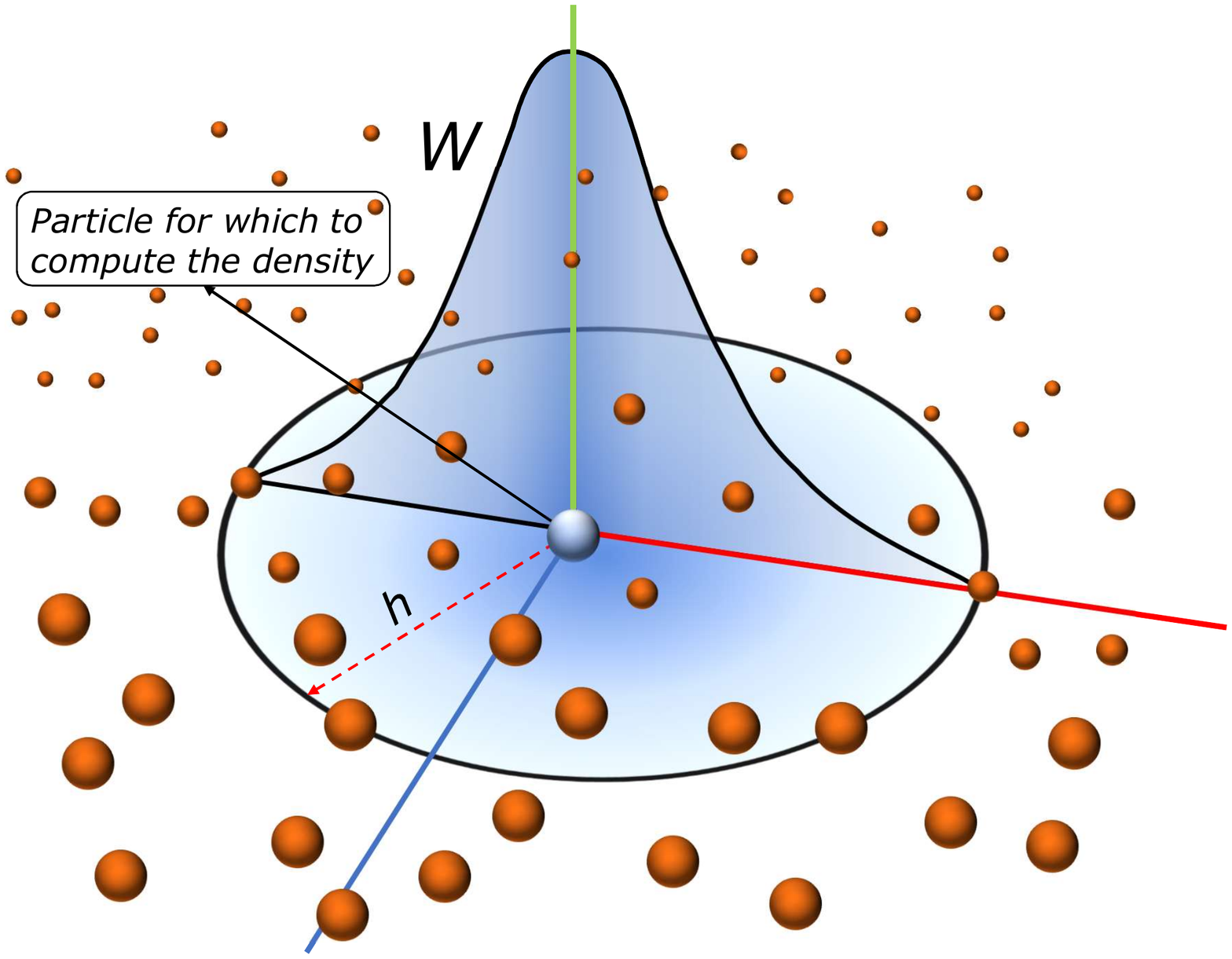}
    \caption{Illustration on how the SPH kernel functions. All particles within radius $h$ 
    contribute a fraction to the density of the target particle. The fraction is related to 
    the distance, the closer the more contribution which is illustrated by the height of the  
    density curve.}
    \label{fig:sph_kernel}
\end{center}
\end{figure}

In SPH the properties of a particle are based on its nearest
neighbours. The contribution of each neighbour depends on the distance
between the particle and the neighbour, where further away neighbours
contribute less than nearby neighbours. For example, the density of a
particle is computed by the sum over neighbouring particles that fall
within the smoothing length distance from the particle. The smoothing
length acts as the search radius and depends on the number of nearby
neighbours. Particles that are located outside this radius do not
contribute to a particle's density. This process is illustrated in
Fig.~\ref{fig:sph_kernel}.  Here the particle that we target is drawn
in the center and around it we have a circle with radius $h$.  The
kernel indicates the strength of the neighbour contribution where from
particles near our target the contribution is higher (peak in the
distribution) and particles that fall outside $h$ contribute
nothing. The exact shape of kernel $W$ depends on the chosen kernel
which we discuss in section~\ref{sect:smoothing_kernels}.

\subsection{Fundamental equations}

In order to accurately discretise the continuous fluid space into
discrete particles each particle has to be associated with a
density. The density of a particle is computed via,

\begin{equation}
\rho_{i}=\sum\limits_{j=1}^{N}{m_jW(|{\bf r}_{ij}|, h_{i})},
\end{equation}
${\bf r}_{ij} = r_{i}-r_{j}$, $W(r,h)$ is the smoothing
kernel and $h_i$ the smoothing length of an individual particle.  The
Smoothing length relates to the particle's density and mass via,

\begin{equation}
h_{i} = h_{\rm fact} n_{i}^{-1/3} = h_{\rm fact}({\frac{m_{i}}{\rho_{i}}})^{1/3}.
\end{equation}
Here $n$ is the particle number density, $\rho$ the
density, $m$ the mass, and $h_{\rm fact}$ a proportional factor that
is specific to the used density kernel~\cite{PRICE2012759}.

In \bonsai\ we implemented the same SPH equations as used by
\phantomx~\cite{phantom_paper} and as such make use of the gradient
based SPH method, we therefore compute the density gradient,

\begin{equation}
\frac{\mathrm{d}\rho_{i}}{\mathrm{d} t} = \frac{1}{\Omega_i}\sum_{N}^{j}m_j(v_i-v_j)\cdot \nabla{i}W_{ij}(h_i)),
\end{equation}
$W_{ij}(h_{i})\equiv W(|r_i - r_j|, h_i)$ is the contribution of the
smoothing kernel between particle $i$ and $j$ given smoothing length
$h_i$.  This is a function of the gradient of the smoothing length,

\begin{equation}
{\Omega_i} \equiv 1 - \frac{\partial h_i }{\partial \rho_i}\sum_{j}^{N}mj\frac{\partial W_{ij}(h_i) }{\partial h_i}.
\end{equation}

\noindent For details on how these equations are derived
see~\cite{doi:10.1046/j.1365-8711.2002.05445.x,
  doi:10.1046/j.1365-8711.2002.05678.x}.

The above equations are independent of the used smoothing kernel, with
the only requirement that it is differentiable.

\subsection{Smoothing kernels}\label{sect:smoothing_kernels}

There are three smoothing kernels implemented in \bonsai. Depending on
the goal of the simulation the user can select, at compile time, which
of these kernels is the most suitable.  The reason for doing this at
compile time is to improve the efficiency, since a number of the
kernel operations involve constants and as such can be optimized by
the compiler.

The following kernels are available:
\begin{itemize}[noitemsep,topsep=0pt]
\item ${\rm M}_4$ cubic spline kernel, a kernel based on the B-spline family~\cite{10.2307/43633538} 
      and the most commonly used SPH kernel~\cite{1985A&A...149..135M}.
\item ${\rm M}_6$ quintic kernel, this is a higher order extension of the ${\rm M}_4$ kernel which requires a larger 
      smoothing range, and therefore more neighbours which makes it computationally more expensive. 
\item Wendland $C^6$, this is one of the more recently developed kernels and proved to be stable against 
      the common pairing instability problem~\cite{2012MNRAS.425.1068D, Wendland1995}.
\end{itemize}

The choice of kernel is an open discussion with no forgone conclusion,
see for example the discussions in ~\cite{PRICE2012759,
  2012MNRAS.425.1068D}.

\subsection{Time-integration}

For the time integration we use the same second order 
Leapfrog integrator~\cite{1995ApJ...443L..93H} as used in \bonsaig.
In this scheme the position
and velocity are predicted to the next simulation time using
previously calculated forces. Then the new densities and forces are
computed after which the velocities are corrected. This is done for
all particles in parallel using the globally determined minimum
time-step\footnote{For this work we make no use of the block-time step
  capabilities of the code.}. This process is described in equations
~\ref{eq:predict_pos} to ~\ref{eq:correct_vel}.
\begin{equation}\label{eq:predict_pos}
r_{1} = r_{0} + v_{0}\delta t + \frac{1}{2}a_{0}(\delta t)^{2},
\end{equation}
\begin{equation}\label{eq:predict_vel}
v_{1p} =  v_{0} + a_{0}\delta t.
\end{equation}
Next the densities and forces are computed after which the velocity
undergoes the correction step,
\begin{equation}\label{eq:correct_vel}
v_{1c} =  v_{1p} + \frac{1}{2}(a_{1} - a_{0})\delta t,
\end{equation}
where $a$ is the acceleration (see Eq.~\ref{eq:acc}).

The time-step is determined after each iteration and constrained by
the Courant time-step~\cite{doi:10.1137/0907039},
\begin{equation}
\label{eq:timestep}
\delta t^{i} \equiv C_{\rm cour} \frac{h_i}{v^{dt}_{sigmax,i}}.
\end{equation}

\noindent We use $C_{\rm cour} = 0.3$ following
~\cite{doi:10.1137/0907039}, and ${v^{dt}_{sigmax,i}}$ is the maximum
signal speed (see Eq.~\ref{eq:vsig}) between particle $i$ and its
neighbours.

The acceleration, $a$, used in
Eqs.~\ref{eq:predict_pos}-\ref{eq:correct_vel} is defined as the sum
over all the contributing neighbours $j$:

\begin{equation}
\label{eq:acc}
a = -\sum_{j} m_{j} \left [ {\frac{P_i + q_{ij}^{i}}{\rho_{i}^{2}\Omega_i}} \nabla_{i}W_{ij}(h_i) + 
{\frac{P_j + q_{ij}^{j}}{\rho_{j}^{2}\Omega_j}} \nabla_{i}W_{ij}(h_j)  \right ] +  {\bm a}_{\rm grav}^{i}.
\end{equation} 

\noindent Here $q_{ij}^{i}$ and $q_{ij}^{j}$ are the artificial
viscosity terms (see below) and ${\bm a}_{\rm grav}^{i}$ is the
Newtonian force exerted on particle $i$. Note that the Newtonian force
is a contribution by all particles and not just the neighbours within
the smoothing range.  In our work the gravitational force is computed
using the Barnes-Hut hierarchical tree algorithm~\cite{1986Natur.324..446B}.

The last property computed during each iteration is the internal
energy which discretised form is given by,

\begin{equation}
\label{eq:energy}
\frac{\mathrm{d}u_i }{\mathrm{d} t}=\frac{{P}_i }{\rho^{2}_{i}\Omega_{i}}
\sum_{j} m_j {\bm v}_{ij}\cdot \nabla_{i}W_{ij}(h_i) + \Lambda_{\rm shock},
\end{equation}

\noindent Here $\Lambda_{\rm shock}$ is the artificial conductivity
shock capturing term, discussed below.

\subsection{Artificial viscosity}

The artificial viscosity terms control the dissipation of the
shock-capturing equations for the equations of
motion~\cite{1997JCoPh.136..298M}.  The switch is defined via,

\begin{equation}
q_{ij}^{i} = 
\begin{cases}
-\frac12 \rho_{i} v_{{\rm sig},i} {\bm v}_{ij}\cdot \hat{\bm{r}}_{ij}, & \bm{v}_{ij}\cdot \hat{\bm{r}}_{ij} < 0. \\
0 & \text{otherwise},
\end{cases}
\end{equation}
with $\hat{\bm{r}}_{ij} \equiv ({\bm{r}}_{i} - {\bm{r}}_{j}) / |
{\bm{r}}_{i} - {\bm{r}}_{j}|$ and ${\bm{v}}_{ij} \equiv {\bm{v}}_{i} -
{\bm{v}}_{j}$ which form the unit vector between the particles and
$v_{\rm sig}$ is the signal speed, given by

\begin{equation}
\label{eq:vsig}
v_{{\rm sig},i} \equiv  \alpha^{\rm AV}c_{{\rm s},i} + \beta^{\rm AV} | {\bm v}_{ij} \cdot \hat{\bm{r}}_{ij} | 
\end{equation}

\noindent where $\alpha^{\rm AV}$ and $\beta^{AV}$ are configuration
parameters that influence how the shocks are treated. In \bonsai\ they
are set at the start of the simulation and are not updated
overtime. This is different from \phantomx\ which uses a more
sophisticated control switch that can update $\alpha^{\rm AV}$, per
particle, during the simulation.  More details and discussions on how
to set these parameters can be found
in~\cite{doi:10.1137/0907039,doi:10.1111/j.1365-2966.2010.16810.x}.

\subsection{Artificial conductivity}
 
As mentioned above to handle shocks in the computation of the internal
energy there is oftentimes a viscosity parameter added. In \bonsai\ we
follow~\cite{phantom_paper} and implemented a conductivity term. The
combination of both the artificial viscosity and artificial
conductivity is defined via,

\begin{equation}
\label{eq:viscosity_shock}
\Lambda_{\rm shock}\equiv -\frac{1}{\Omega_{i}}
\sum_{j} m_{j}v_{{\rm sig},i}
     \frac{1}{2}({\bm v}_{ij}\cdot\hat{{\bm r}}_{ij})^2 F_{ij}(h_i) 
    + \sum_{j} m_{j}\alpha_{u}v^{u}_{\rm sig}(u_i - u_j)
\frac{F_{ij}(h_i)}{\Omega_{i}\rho{i}} + 
\frac{F_{ij}(h_j)}{\Omega_{j}\rho{j}} 
\end{equation}

\noindent Here $\alpha_{u}$ is the configurable parameter that
controls the strength of the thermal conductivity. In general, we keep
this fixed to $\alpha_{u}=1$.  The force $F_{ij}$ is defined via,
\begin{equation}
F_{ij} \equiv \frac{C_{\rm norm}}{h^4}f'(q).
\end{equation}

\noindent Here $C_{\rm norm}$ is a property of the chosen smoothing
kernel and $f'(q)$ the derivative of $q = (|r_{i}-r_{j}|/h)$.

Finally, the signal speed, $v^u_{\rm sig}$, is defined via 
\begin{equation}
v^{u}_{\rm sig} = \sqrt{\frac{|P_i - P_j|}{\bar{\rho}_{ij}}},
\end{equation}

\noindent where $P$ is the pressure of a particle and 
$\bar{\rho}_{ij}$ the average density of particles $i$ and $j$.

\section{Implementation}\label{Sect:Implementation}

This section starts with a short overview of \bonsaig\, followed by a
short description of the most important algorithms and how they are 
implemented on the GPU. For each of these algorithm we describe the 
differences between \bonsaig\ and \bonsai\ to indicate how 
adding support for fluid dynamics affects the basic tree related algorithms. 
A full description on how \bonsaig\ works can be found in
\cite{2012JCoPh.231.2825B,Bedorf:2014:PGT:2683593.2683600} or by
inspecting the source-code in the online repository~\cite{Bonsai}.

The development of \bonsaig\ started in 2010 with the goal to develop
an $N$-body code that was, from the ground up, optimized for GPU
accelerators. When the code was completed the full software stack was
executed on the GPU.  This eliminated data transfer requirements and
allowed the $\mathcal{O}(N)$ parts of the code to take advantage of
the additional processing and bandwidth capabilities of the GPU.  When
targeting a single GPU, the CPU's tasks are limited to basic
operations and orchestrating compute work on the GPU.  This frees up
the CPU cores for other work, such as on-the-fly post-processing or
visualizations. When executing \bonsaig\ in a distributed setup the CPU
is responsible for handling (network) communication between the GPUs.
The communication patterns depend on the number of nodes and the
simulated model, exactly the kind of irregular tasks for which the CPU
is perfectly suited. The distribution of work between the CPU and
GPU, is such that each can focus on its strengths, and allows 
\bonsaig\ to scale to thousands of GPUs while maintaining high 
computational efficiency.

For the SPH additions we stick to the above design pattern and keep
all the compute work and data storage on the GPU. In practice this
means that additional memory is reserved for storing the fluid dynamic
properties such as pressure, density, energy, etc. Furthermore,
compute kernels are added to compute hydro properties.


The GPU portions of the software are developed using {\tt CUDA} which in practice 
means that only NVIDIA GPU hardware is supported.
\bonsai\ works on the Tesla, GeForce and Quadro GPU series as long 
as the compute capability of the device is 3.0 or greater.

\subsection{Tree Construction}

The tree-code algorithm is based on the assumption that particles can be grouped in a 
hierarchical data-structure. Most CPU based algorithms create an octree data structure by performing 
sequential particle insertion which adds particles to a box that encloses the spatial coordinates 
of the particles. Once the box is full the box is split up into 8 sub-boxes (hence the name octree)
and the particles that were in the original box are divided over those sub-boxes. For GPU based 
algorithms this is not efficient as there would be too many race conditions when multiple 
threads become involved. 
Therefore, we use a different kind of method that can be 
executed by many threads in parallel. The method uses a space filling curve~\cite{citeulike:2861104}
to order particles into the boxes. Each particle is assigned a unique location on the curve, based 
on the coordinates of the particle. Next, the particles are sorted such that their order in memory 
follows the space filling curve. Both these operations are executed on the GPU,
where Thrust\footnote{https://developer.nvidia.com/thrust} is used to perform the sort operation. Next the octree is constructed by 
chopping the space filling curve into sections where each section refers to part of the tree.
This way the tree is built level by level until the smallest section of non-chopped curve contains 
at most $N_{\rm leaf}$ particles. Where $N_{\rm leaf}$ stands for the maximum number of particles
that is assigned to a leaf, an end point of a tree branch.

Using the same set of sorted particle we create \emph{particle groups}. These particle groups are used 
during the tree-traverse, where a group of particles traverses the tree instead of individual particles. For the 
group construction we again chop the space filling curve into sections. The sections are chopped 
into smaller sections until each section contains at most $N_{\rm group}$ particles. 

The above described method is the same for both \bonsaig\ and \bonsai\ with the difference that 
for \bonsai\ lower values are used for $N_{\rm leaf}$ and $N_{\rm group}$. The lower values give 
better performance during the tree-traverse required for computing SPH properties
as discussed in the next section.

\subsection{Tree-traverse for SPH}

The tree-traverse for the density/hydro-force computations are similar
to the method used for the gravity computation. However, where the
gravitational force requires information from all particles, either
via direct interaction or via multipole expansion approximations, the
SPH method only requires information from particles that fall within
the smoothing range (see Fig.~\ref{fig:sph_kernel}).
This has a number of consequences which we will list after giving a 
global description of the tree-traverse method.

CPU based tree-traverse algorithms are often implemented via a recursive algorithm. 
For the GPU processor this is not a good fit and instead we use a distributed breath 
first traversal algorithm. Furthermore, particles do not traverse the tree individually 
but in a group of particles, this is known as Barnes' vectorization~\cite{1990JCoPh..87..161B}
and improves the GPU utilization as groups of particles perform the same operations in parallel.

For our GPU implementation we make extensive use of in- and exclusive scan algorithms, for example
to expand 
compressed node indices. Say, if a leaf contains 8 particles then we only store the 
index of the first particle, and the number of particles. This tuple, for example (104, 8), is then 
expanded as follows: 104, 105, 106, 107, 108, 109, 110, 111.
To optimize the performance and reduce the amount of memory resources required, the scan 
algorithms are implemented with the use of shuffle instructions and embedded PTX code.

During the tree-traverse individual tree-nodes are tested and for each node is decided if it has to 
be expanded (traversed further) or that it falls outside of the search range. The possible options are,

\begin{itemize}[noitemsep,topsep=0pt]
\item If a node falls outside the search range then it is either discarded (when computing fluid dynamic 
properties), or it is put on a \emph{multipole approximation evaluation list} (gravity computation). 
The list is stored in the GPUs on chip shared memory and therefore has a relative limited size,
but allows for quick access.
Once the list is full it is processed in parallel by the threads traversing the tree and the 
multipole approximation between the tree-nodes and the particles that are part 
of the group traversing the tree is computed. 
\item If a node falls inside the search range and it is not a leaf then it will be added to the \emph{next level list}
and processed further during the next loop of the tree-traverse algorithm.
\item If a node falls inside the search range and it is a leaf then the individual particles 
of the leaf are added to the \emph{particle evaluation list}. Once the list is full 
the list is processed. This is described in more detail in Sect.~\ref{sect:int_list_proc}.
\end{itemize}

The tree-traverse is continued until the next level list is empty which indicates that all 
relevant sections of the tree have been processed. 


The useful or not decision as made during the tree-traverse is different for the gravity 
and fluid dynamic computations. For gravity the decision is based on the distance 
between a particles group and a tree-node with respect to a specified \emph{multipole acceptance 
criteria}. While for fluid dynamics it is based on the distance between a particle group 
and the tree-node and if this is less than the smoothing range of the particle group. 
\\
This has the following consequences for \bonsai\:
\begin{itemize}[noitemsep,topsep=5pt]
\item The difference between interaction lists of individual particles
  is larger when compared to the gravity interaction lists. A too large 
  a difference leads to executing unneeded computations. To reduce 
  the difference we use a more fine-grained interaction path, achieved 
  by using smaller values for $N_{\rm leaf}$ and $N_{\rm group}$ 
  when computing fluid properties. This causes particle groups and leaf nodes to be 
  physically smaller, and thereby reducing the number of non-useful interactions.  
\item For SPH there are two properties that are computed via tree-traverse 
  operations, namely the density and the hydrodynamic force which both
  require a slightly modified tree-traverse. The difference lies in how it is 
  decided which particles should interact. For the density computation this is determined
  via the smoothing range of the particle traversing the tree. For
  the hydro-force computation it is required, in order to preserve
  angular momentum, that a force is computed if one particle falls within
  the smoothing range of the other. Therefore we use the
  maximum smoothing of the two candidate particles to determine if the
  particle has to be added to the interaction list.
\end{itemize}

The lower values for $N_{\rm leaf}$ and $N_{\rm group}$ reduce the 
efficiency of the gravity computation, but this is offset by large
efficiency gains for the density and hydro-force computations.

\subsubsection{Interaction list processing}
\label{sect:int_list_proc}
When processing the interaction list for the fluid dynamic computations,
we either execute density or
hydro-force computations.  The function to execute is a templatized
parameter of the tree-traverse code.  Each thread of the group is
responsible for loading part of the interaction list in memory and
sharing it with neighbouring threads when the density and hydro-force
functions are executed.  After processing the list the partial results
are returned and the tree-traverse continues.

As with the gravity computation these functions are optimized and make
use of the shared memory and shuffle instructions available on NVIDIA
GPUs via the CUDA language.  Compared to the gravity computation the
number of required resources is considerable larger when computing SPH
related properties. This has a negative effect on the performance,
because data will be flushed from registers to main memory. However,
the overall performance is still better than when using CPUs (see
Sect.~\ref{Sect:Results}).

During each time-step we run the density computation multiple times in
order to let it converge on the correct number of
neighbours. Currently this iteration is done 3 times for all
particles, a future optimization would be to make the number of
iterations dynamic with a per-particle group coarseness. The
hydro-force computation is executed only once as there is no
convergence requirement.

\subsection{Particle integration}

In order to move particles forward in time and update properties such as density and energy the 
newly computed forces have to be applied on the particles. For this a grid of compute threads 
is launched on the GPU. Each thread is responsible for processing a single particle. This method 
is the same for both the gravity and hydrodynamic version of the software.

\subsection{Multi-GPU}

The multi-GPU implementation is an extension of the gravity version,
described in detail in~\cite{Bedorf:2014:PGT:2683593.2683600}, 
which uses the local essential tree (LET) method 
to exchange data with neighbouring processes~\cite{Warren:1992:ANS:147877.148090}. 
In \bonsaig\ the LET tree is built on the CPU, while the GPU is computing 
gravitational forces. The CPU then sends/receives the LET trees that contain 
the particle properties required to compute the total gravitational
force exerted on a particle. 
For \bonsai\ the method is extended to, next to the properties required for gravity, 
include the tree-node and particle properties required for hydrodynamics.
In addition we improved some of the existing pre-compute operations (e.g. detection of 
domain boundaries) by migrating them from the CPU to the GPU in order to reduce the 
amount of required memory copies.

In practice it turned out that the parallelization method did not work as
efficiently for SPH as it does for gravity. The tree-traverse to create LET 
structures has to perform a deeper traversal to determine which particles are
important. This is a consequence of having more fine grained 
groups (8 vs 64 particles per group). In practice this leads to 
increased CPU processing time which hinders the scalability and 
results in nearly no improvement in execution time.
It is possible to tune this selection process further,
either by setting a depth limit on the traversal and instead send more
data than required, but it is unlikely to be sufficient. 
We expect that to make the selection more efficient, at a minimum, we would
have to use a different domain decomposition method.  Adopting the
orthogonal bisection method~\cite{Ishiyama:2012:PAN:2388996.2389003},
instead of the Peano-Hilbert curve, would allow us to significantly speed 
up the selection of particles that are within the search radius of the domain
boundaries. This would result in a considerable speed up for the
multi-GPU implementation of the code as the CPU will be less constrained.
In this work we focus on single GPU performance and correctness and therefore leave 
this multi-GPU improvement for a future version.

%
%
%
%

\section{Results}\label{Sect:Results}

To validate the code and show the conservation properties we use a
number of well-known SPH tests. We compare our results with the
analytic solutions as well as with the results of
\phantomx\ ~\cite{phantom_paper}. For all tests we use the
\phantomx\ setup programs to generate the initial conditions.

Because \bonsai\ only support a global, constant, artificial viscosity
we changed the default settings of \phantomx\ to match this. For some
of the tests this lead to quantitatively slightly different results
when compared to those published in~\cite{phantom_paper}.  Unless
indicated otherwise we use the ${\rm M}_4$ cubic spline kernel, the
Courant time-step parameter $C_{\rm cour}=0.3$, an adiabatic index of
$\gamma={\frac{5}{3}}$, and the viscosity switches $\alpha^{\rm AV}=1$,
$\beta^{AV}=2$ and $\alpha_{u}=1$.  Following the standard SPH
validation tests we further show the scaling performance and energy
conservation properties of \bonsai.

We used {\tt SPLASH} ~\cite{2007PASA...24..159P} for plotting and
extracting the exact solutions where applicable.

The computing system we used is an IBM S822LC (Minsky) system. This
machine has two Power8 CPUs and 4 NVIDIA Tesla P100 GPUs. The Power8
CPUs have 8 cores, and each core can handle 8 threads. This gives a
total of 128 threads that can be concurrently active. The operating
system used is Red Hat 7.3, combined with CUDA 9.1.

\subsection{Sod Shock Tube}

Our first test is the standard Sod shock tube
test~\cite{1978JCoPh..27....1S}. Here we configure two different
fluid states (left and right) with an initial discontinuity between
the two states at $x=0$. The left state ($x \leq 0$) has $[\rho,P] =
[1,1]$ with $256\times24\times24$ particles, while for the right state
($x > 0$) we have $[\rho, P] = [0.125, 0.1]$ with
$128\times12\times12$ particles. For this test the ${\rm M}_6$ quintic
spline kernel is used (rather than the standard ${\rm M}_4$ setting)
and periodic boundaries for the $y$ and $z$ axes.  Details on how the
3D initial conditions are generated can be found
in~\cite{phantom_paper}.

\begin{figure*}
    \includegraphics[width=\columnwidth]{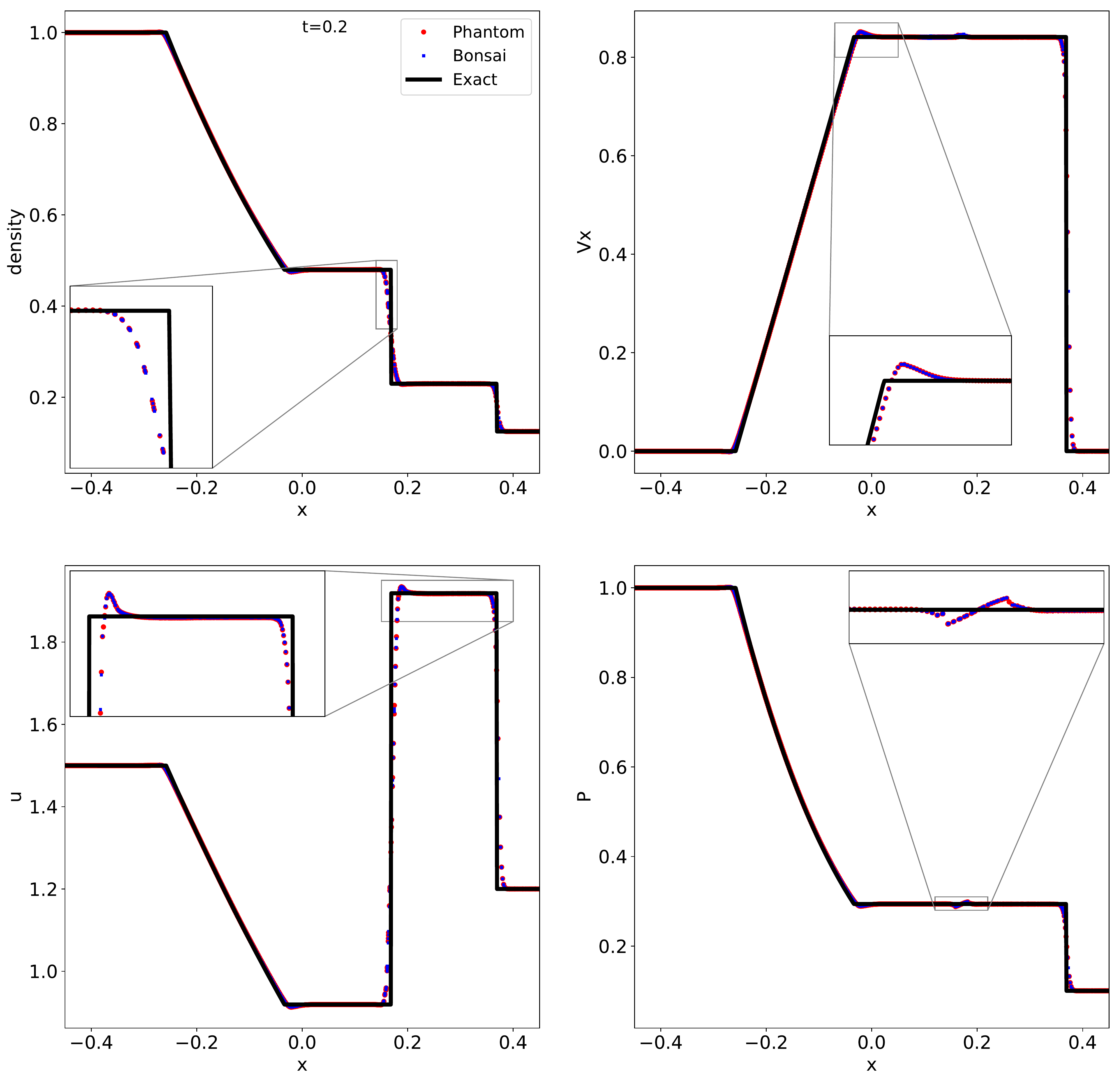}
    \caption{Sod shock test. Plotted are \bonsai\ (blue dots),
      \phantomx\ (red dots), and the exact solution (black). From the
      top left to bottom right the panels show the density, velocity
      in the $x$-direction, energy and pressure plotted against the
      $x$-position.  The results are shown for $t=0.2$.}
    \label{fig:sod_shock}
\end{figure*}

In Fig.~\ref{fig:sod_shock} we present the results at $t=0.1$ for
\bonsai, \phantomx\ and the exact solution. The \bonsai\ and
\phantomx\ results are qualitatively indistinguishable.

\subsection{Blast wave}

\begin{figure*}
    \includegraphics[width=\columnwidth]{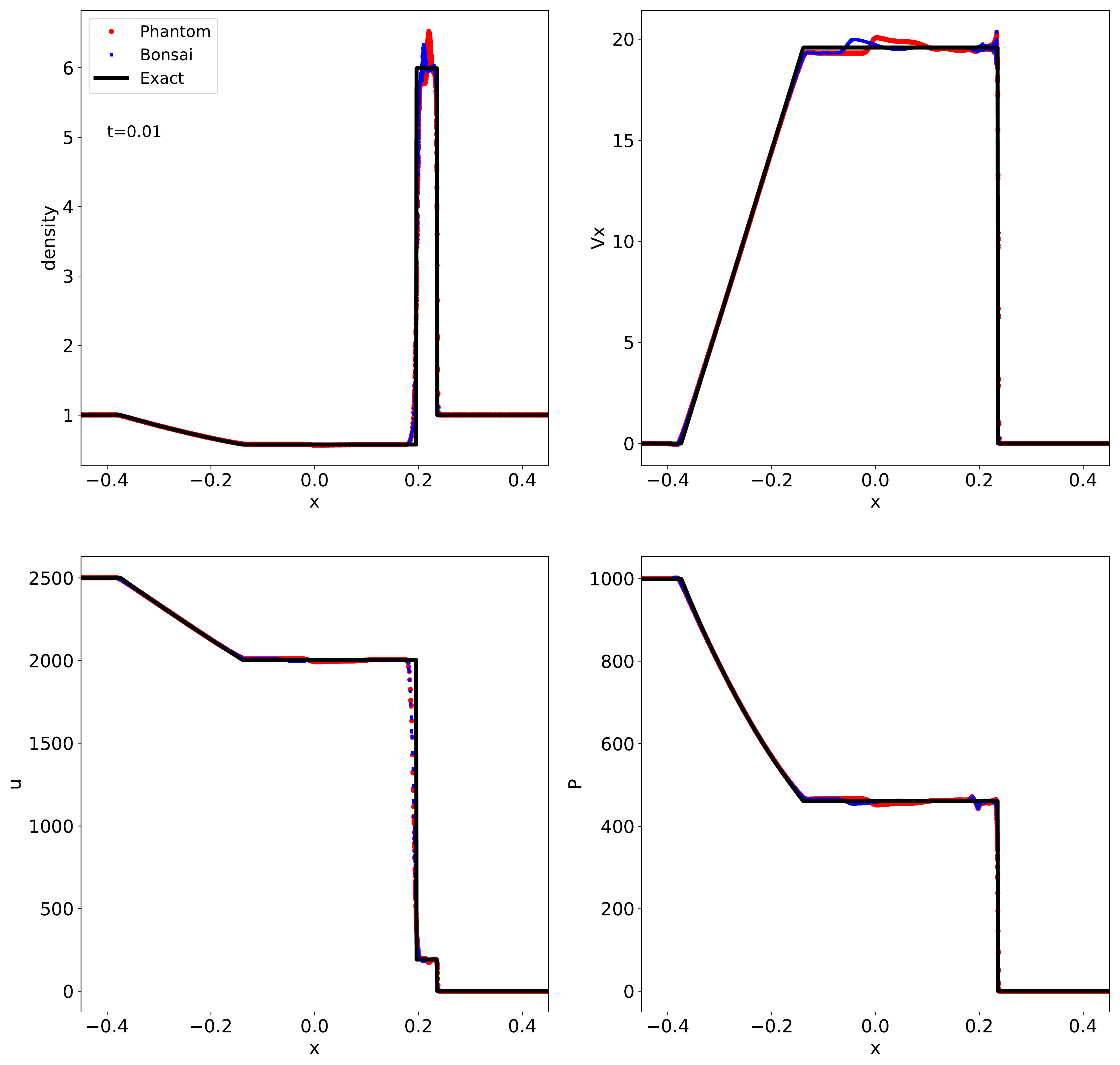}
    \caption{Blast wave test. Plotted are \bonsai\ (blue dots), \phantomx\
      (red dots), and the exact solution (black). From the top left to
      bottom right the panels show the density, velocity in the
      $x$-direction, energy and pressure plotted against the
      $x$-position. The results are shown for $t=0.01$.}
    \label{fig:blast_wave}
\end{figure*}

As a second test we perform the blast wave test~\cite{1997JCoPh.136..298M}, 
which is more sensitive to
implementation details as the differences between the left and right
states is much larger.  Here $[\rho,P] = [1,1000]$ for $x \leq 0$ with
$400\times12\times12$ particles and $[\rho, P] = [1.0, 0.1]$ with
$400\times12\times2$ particles for $x > 0$.  For this simulation we
set $\gamma=\frac{7}{5}$, while the same viscosity and kernel settings
as with the sod shock tube test are used. The results are presented in
Fig.~\ref{fig:blast_wave}, where the exact solution is indicated with
the solid line and the \bonsai\ and \phantomx\ results are presented
with the symbols.  As with Fig.~\ref{fig:sod_shock} the results
between the exact solution and those of the simulations are comparable
with the exception of small quantitative differences in the density
and velocity profiles along the contact discontinuity at $x=0.21$.
There appears to be a minor phase difference between \phantomx\ and
\bonsai, but both codes show similar behaviour and the same error
range. The phase difference is caused by the time-stepping method, but
when we adopt a smaller value for $C_{cour}$ the phase difference is
reduced.

\subsection{Sedov blast wave}

\begin{figure}
\begin{center}
    \includegraphics[width=0.5\columnwidth]{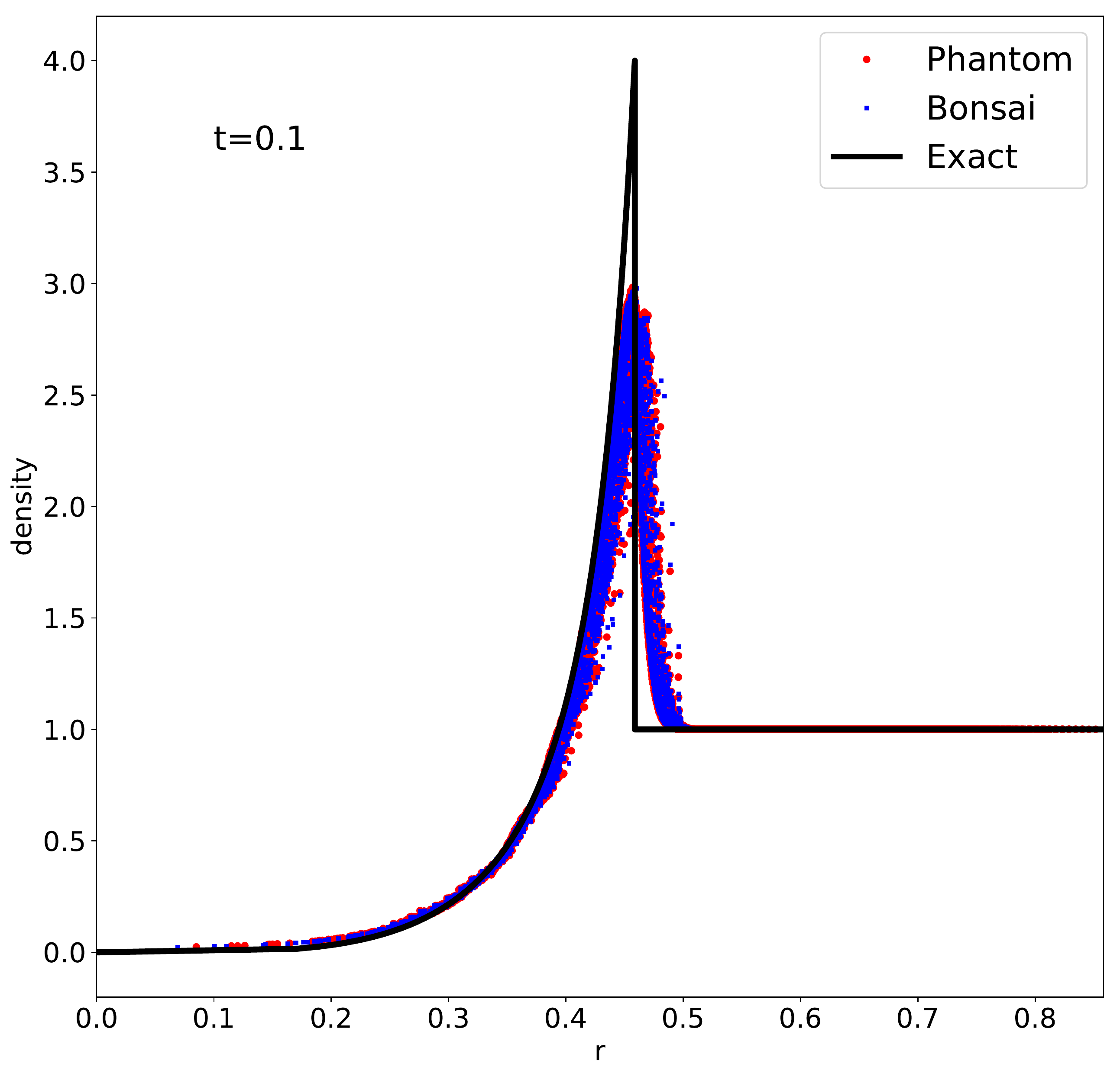}
    \caption{Sedov-Taylor Blast wave.  Plotted are \bonsai\ (blue
      dots), \phantomx\ (red dots), and the exact solution (solid
      line).  The results are shown for $t=0.1$.  }
    \label{fig:sedov}
\end{center}    
\end{figure}

The Sedov-Taylor blast-wave test~\cite{1959sdmm.book.....S} can be
compared to an analytic solution. This test follows the propagation of
a blast wave in a spherical medium, and is often used to estimate the
effect of supernovae explosions.  We configure a uniform 3D box in
which we place, at the center, a sphere composed of $100^3$ particles.
The center particles are given a high initial energy which causes the
explosive blast once the simulation is started.  The results are
presented in Fig.~\ref{fig:sedov} and show \bonsai, \phantomx\ and the
analytic solution.  Both the simulation codes fail to resolve the peak
density that is predicted from the analytic solution, but other than
that the results are consistent with the prediction.

\subsection{Kelvin-Helmholtz instability}


The Kelvin-Helmholtz (KH) instability test demonstrates the mixing
behavior of two fluids with different densities at the moment the
instability sets
in~\cite{doi:10.1111/j.1365-2966.2007.12183.x,2008JCoPh.22710040P}. Much
has been written about this test in particular with respect to the
differences between SPH and grid codes. Traditionally SPH codes were
unable to properly resolve this instability, but the addition of
artificial viscosity and conductivity helped to resolve
this~\cite{PRICE2012759}. Furthermore, the initial conditions have to
be generated properly for a fair comparison between the various
methods to simulate fluid
dynamics~\cite{doi:10.1111/j.1365-2966.2009.15823.x}. In this work we
use the method described in
~\cite{doi:10.1111/j.1365-2966.2009.15823.x} which is implemented in
the initial conditions generator of \cite{phantom_paper}. The KH test
is in two dimensions but given that the code operates using
three-dimensional coordinates it is executed as a flat bar. The box
coordinates are between 0 and 1 in the $x$ and $y$ direction. Details
on generating the initial conditions can be found in (\cite{phantom_paper}, 
section 5.1.4).  The adiabatic index of the simulated fluid
is $\gamma=\frac{5}{3}$, we use the ${\rm M}_4$ cubic spline kernel and
the default settings for the viscosity switches.

Since there is no analytic solution, the only way to validate our
results (apart from energy conservation tests) is to compare it with
previous implementations. We therefore ran the same initial conditions
with \phantomx, where we configured the free parameters to match those
of \bonsai. For all the figures in this section we show the cross
section of the particle density at $z=0$.

\begin{figure*}
    \includegraphics[width=\columnwidth]{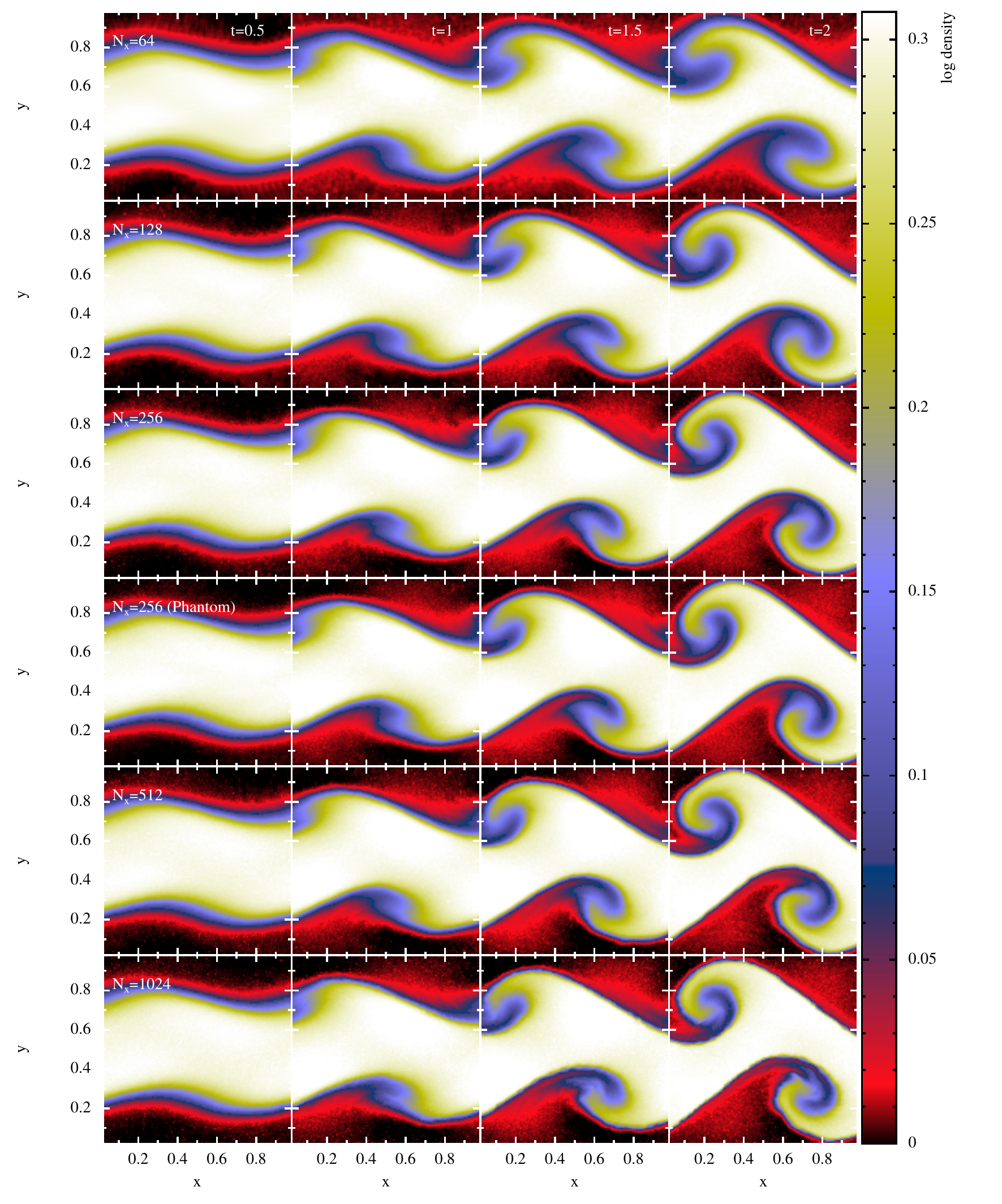}
    \caption{Kelvin-Helmholtz instability test. Presented are 5
      different resolutions ($N_x$=64, 128, 256, 512 and 1024) where
      each column presents a different time-step ($t=$ 0.5, 1.0, 1.5
      and 2.0).  The first 3 and final 2 rows are results from
      \bonsai\ the fourth row is generated using \phantomx. The
      visualizations are the cross sections at $z=0$.
    \label{fig:KH_test1}
    }
\end{figure*}

In Fig.~\ref{fig:KH_test1} we present a similar figure as presented
in~\cite{doi:10.1111/j.1365-2966.2009.15823.x, phantom_paper} which
shows the development of the instabilities as computed using
\bonsai\ for 5 different resolutions ($n_x$=64, 128, 256, 512 and 1024)
between $t=0.5$ and $t=2$.  In between the \bonsai\ results we show
the $n_x$=256 result as computed using \phantomx. The results are
qualitatively indistinguishable and both codes show the same behavior
until the end of the simulation at $t=10$ (not shown here).

\begin{figure}
\begin{center}
    \includegraphics[width=0.7\columnwidth]{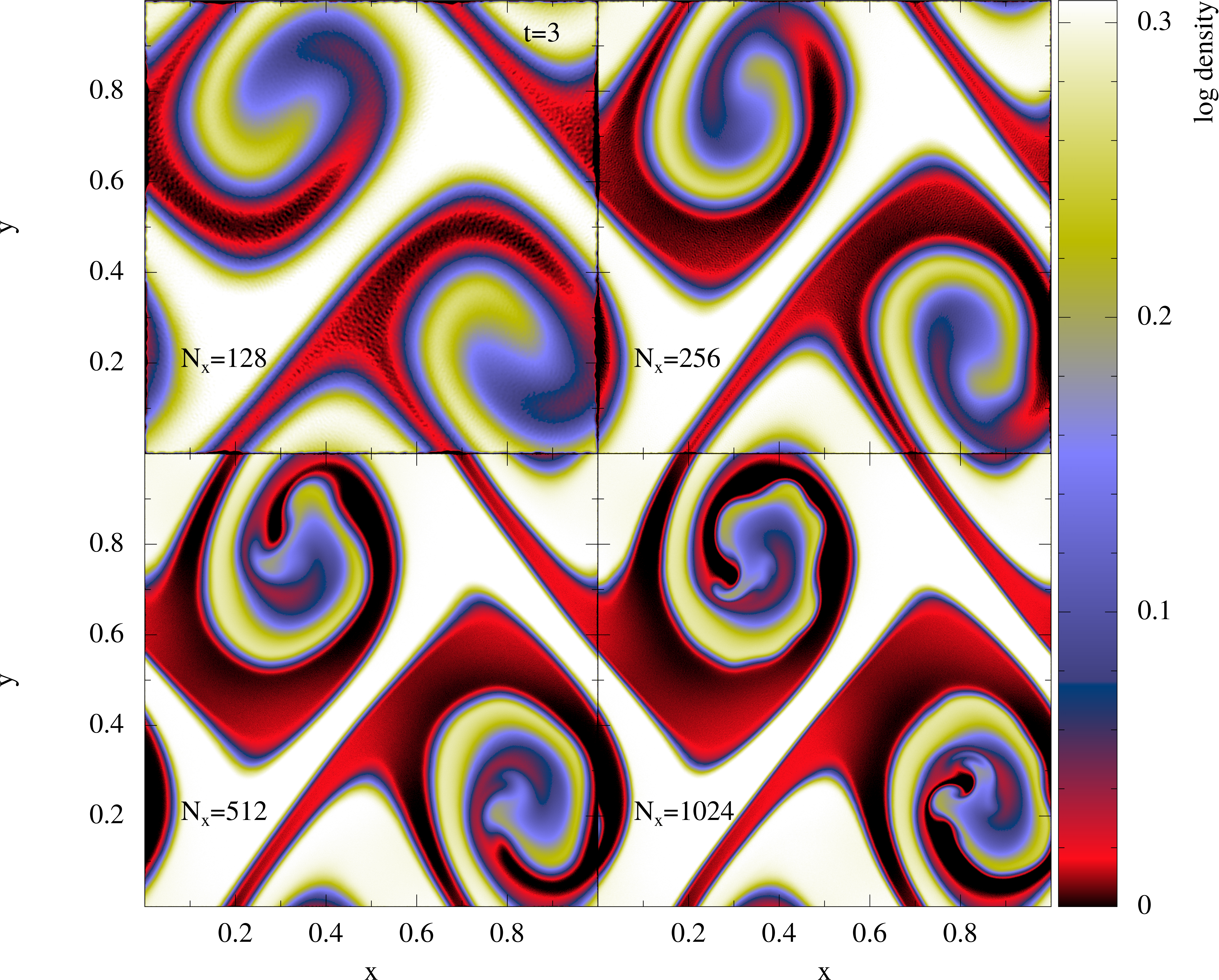}
    \caption{Kelvin-Helmholtz instability test. Presented are 4
      different resolutions ($N_x$=128, 256, 512 and 1024) where each
      resolution occupies one of the panels. Results are generated
      using \bonsai\ with the ${\rm M}_6$ quintic kernel.  Note this
      is at $t=3$ while Fig.~\ref{fig:KH_test1} is at $t=2$.}
    \label{fig:KH_test2}
\end{center}    
\end{figure}

At higher resolutions we see some noise appearing in the cross
sections and the highest density contrasts become somewhat fuzzy. We
therefore repeat the simulations using the higher order ${\rm M}_6$
quintic kernel. The higher order kernel should give smoother results
because more neighbours are involved. This time we used 4 different
resolutions ($n_x$=128, 256, 512 and 1024) and compared the $t=3$
snapshot. The results are presented in Fig.~\ref{fig:KH_test2}, in a
manner similar to the results from {\tt ENZO} presented
in~\cite{doi:10.1111/j.1365-2966.2009.15823.x}[Fig. 8].  Both our
figures use the same colour scale.

The results are much smoother than the results from
Fig.~\ref{fig:KH_test1} for $n_x$=512 and 1024.  This demonstrates
that \bonsai\ behaves as expected and is able to resolve the tiny
features required to generate mixing.

\subsection{Energy conservation}

In the previous sections we used the analytic solution to validate the
code, alternatively it is also possible to keep track of the energy
conservation to verify that the code behaves correctly.

We selected three simulations from the previous sections and extracted
the energy conservation over the course of the simulation for both
\phantomx\ and \bonsai.  We selected the \emph{Sod Shock Tube}, the
\emph{Sedov blast Wave}, and the \emph{Kelvin-Helmholtz instability}
run with $N_x$=256 using the $M_6$ kernel.

\begin{figure*}
    \includegraphics[width=1.0\columnwidth]{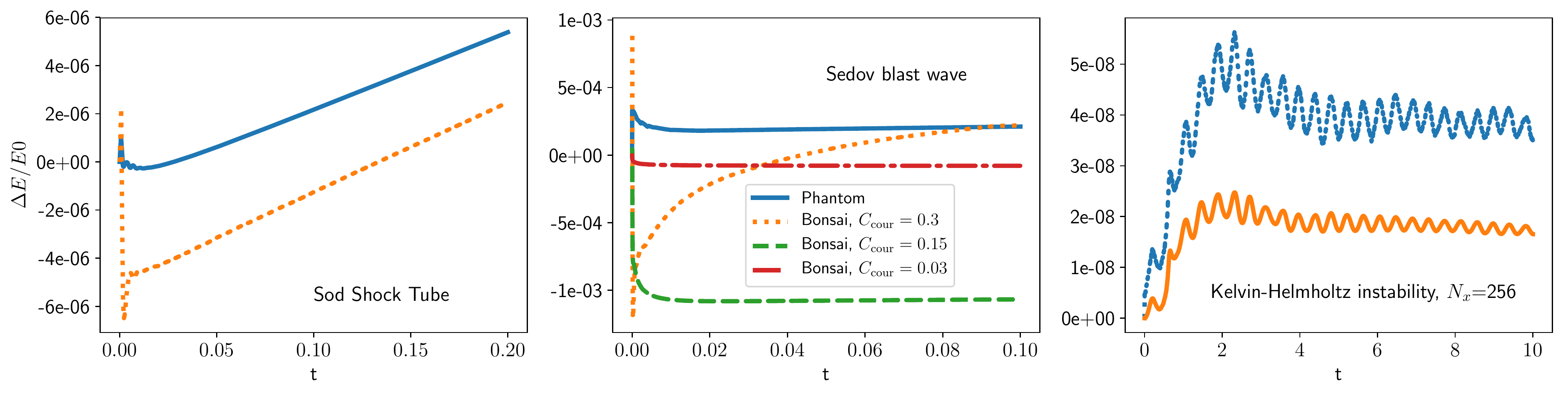}
    \caption{Energy error of \bonsai\ vs \phantomx\ for the \emph{Sod
        Shock Tube} (left panel), \emph{Sedov blast Wave} (middle
      panel) and \emph{Kelvin-Helmholtz instability} (right panel).
      In all panels the $x$-axis indicates the time since the start of
      the simulation and the $y$-axis the energy error via
      Eq.~\ref{eq:energy_error}.  For all panels, the solid line shows the
      results as obtained with \phantomx\ and the dotted line the
      results of \bonsai\ using the default time-step. In addition,
      the middle panel shows the results of \bonsai\ using $C_{\rm
        cour}$=0.15 (dashed-line) and $C_{\rm cour}$=0.03 (dash-dotted
      line).
    \label{fig:energy_error}}
\end{figure*}

The comparison is presented in Fig.~\ref{fig:energy_error}, and
computed using,
\begin{equation}
\label{eq:energy_error}
dE = (E_0 -E_t) / E_0
\end{equation}

In the left panel the results of the \emph{Sod Shock Tube} are
presented and both \bonsai\ and \phantomx\ show similar behaviour
where both codes give an energy error on the order of $10^{-6}$.

The middle panel shows the \emph{Sedov blast Wave} test. Using the
default time-step parameter we found that the energy error of
\bonsai\ in the first few steps behaves quite erratically, we
therefore ran two more tests where we decreased the $C_{\rm cour}$
value (Eq.~\ref{eq:timestep}). This stabilized the results, and
brought it more in line with the results of \phantomx\ which, in
contrast to \bonsai\, uses a combination of 6 different time-step
criteria to determine the step being used.

The right panel shows the \emph{Kelvin-Helmholtz instability}
data. The result of \bonsai\ is slightly worse than that of \phantomx,
but is within the same order of magnitude and shows similar behaviour.

The main reason, apart from the time-step method mentioned above, for
the difference between the two codes is the used numerical
precision. \bonsai\ uses {\tt float32} whereas \phantomx\ uses the
{\tt float64} data-type. This higher precision improves the accuracy
and reduces the noise in the computations.

\subsection{Performance}

One of our goals of developing \bonsai\ was to get access to a faster
SPH code by taking advantage of the GPU's computational resources. In
this subsection we therefore compare the performance of \bonsai\ with
that of \phantomx. For \bonsai\ we used a single P100 GPU and two CPU
threads. We use one thread for controlling the GPU and the other
thread for writing data, no further threads are required as we do not
use multiple GPUs, nor do we do any post-processing. We used the
following properties for the tree-structure, {$N_{\rm leaf}=16$,
  $N_{\rm crit}=8$, no further tuning is required to run
  \bonsai\footnote{for details see~\cite{2012JCoPh.231.2825B}}.

As mentioned at the beginning of this section the used {\tt Power8}
CPU is capable of running 64 threads per CPU. These threads, however,
do share some of the hardware resources and therefore the ideal number
of threads differs per application. Given that \phantomx\ is capable
of making optimal use of multiple threads~\cite{phantom_paper} we
had to find the most optimal number of threads to make a fair
comparison between both codes. For this we ran a set of Kelvin
Helmholtz test calculations to determine the optimum number of CPU
threads. The results of this experiment is presented in
Fig.~\ref{fig:phantom_threads}. The speed-up indicates that it is
beneficial to add additional CPU threads, until the peak is reached
for $N_{\rm thread}=64$. Using more than 64 threads does not lead to
better performance because the threads are competing for the same
resources. We therefore set the number of CPU threads used by
\phantomx\ to 64.

\begin{figure}
\begin{center}
    \includegraphics[width=0.5\columnwidth]{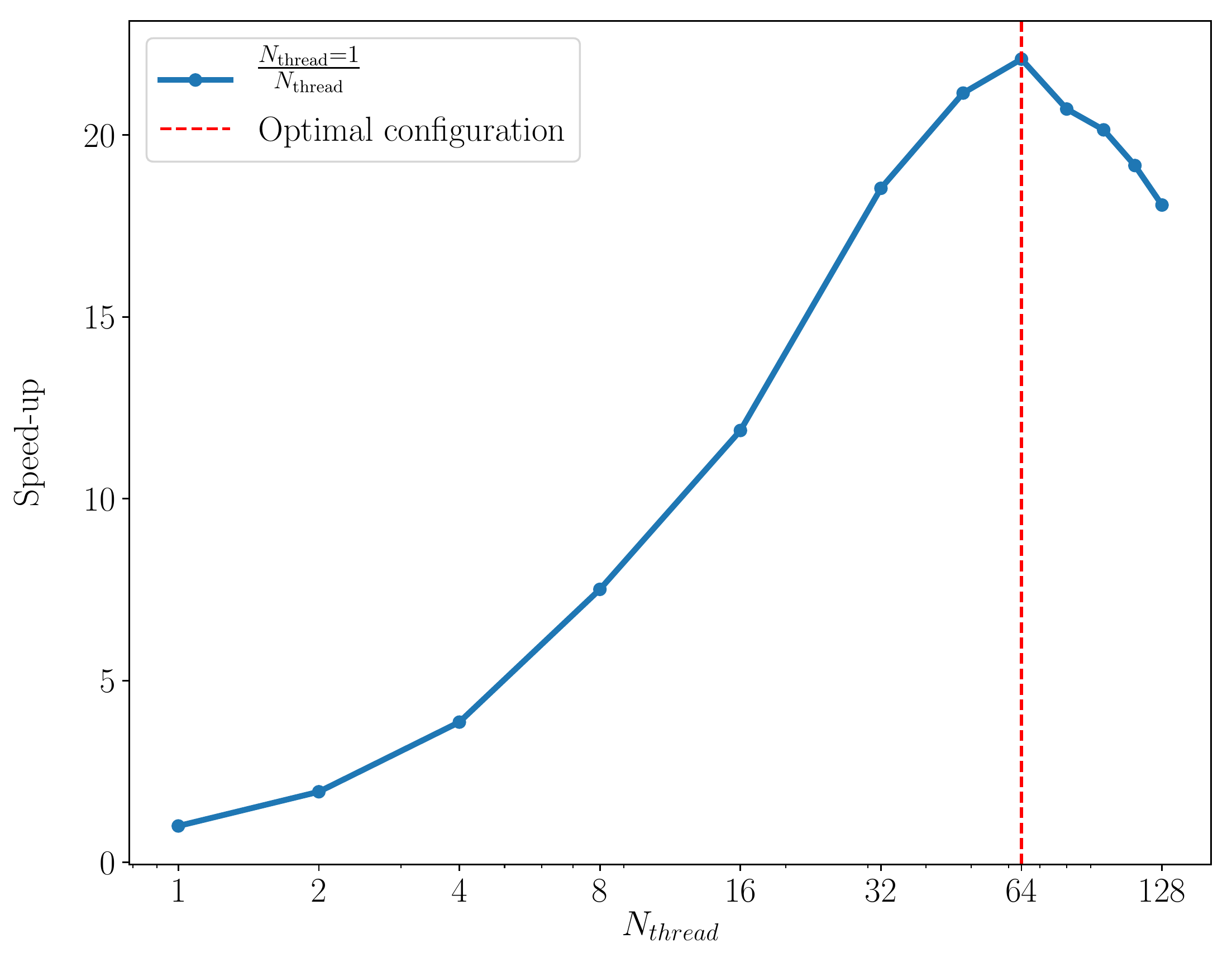}
    \caption{Effect of increasing the number of compute threads used
      by \phantomx\ on the execution speed.  The $x$-axis shows the
      number of {\tt OpenMP} threads and the $y$-axis the speed-up
      compared to single thread execution. The relative speed is
      indicated by the solid blue line, the best performance is
      reached when $N_{\rm thread}$=64, indicated by the vertical red
      dashed line.  Data generated using the $N_{x}$=128
      Kelvin-Helmholtz dataset.}
    \label{fig:phantom_threads}
\end{center}    
\end{figure}

Next, we compare the performance of \bonsai\ with respect to
\phantomx. For this we use the KH simulations. We chose this test because,
of all our models these simulations contained the most particles and has a wide range of density
contrasts. For the reasons described above we use $N_{\rm thread}$=64
for \phantomx\ and compare that to the single GPU timing results of
\bonsai. 

Because \bonsai\ performs fewer time-steps per simulation than \phantomx\ we base 
our performance comparison on the average wall-clock time \emph{per simulation step} instead of the total 
wall-clock time. To ensure that reported numbers are stable, and based on enough data points, 
we execute the KH simulation for 10 time-units (2 units for $N_{512}$ and $N_{1024}$). This results
in thousands of time-steps per simulation. We verified that the time-per-step is roughly constant 
over the course of the simulation to ensure that the reported comparison is valid for the 
whole simulation. We furthermore performed multiple independent simulations for the $N_{64}$ 
configuration to verify that the timing data between runs is consistent. In all cases we found that 
there is little to no variation between runs and over the course of the run so we only 
present the average data of a single run. This also allowed us to simulate a shorter time-frame 
for the large $N$ models in order to get the results within a day instead of a month. 

In order to make a proper performance comparison between \bonsai\ and \phantomx\ 
we have to take into account the numerical precision difference we mentioned earlier. 
Therefore we did the performance evaluation using two different versions of \phantomx. 
The first version uses the default compiler settings which results in double precision (64bit) 
floating point operations. For the second version we modified the compiler 
flags\footnote{specifically the {\tt "DOUBLEPRECISION"} flag, used to disable 64bit computations.}
to build a version that only uses 32bit floating point operations, e.g. the same accuracy 
as \bonsai. 

To compute the speed-up we divide the averaged time per simulation step of \phantomx\ with 
that of \bonsai\, the results are presented in
Fig.~\ref{fig:KH_performance}. We see that \bonsai\ is a factor 4 to
10 times faster than \phantomx. Where the speed-up is smaller for
smaller datasets, which can easily be explained by the fact that for
smaller dataset sizes the GPU is underutilized.  For our largest
dataset size (14M particles) \bonsai\ is almost a factor 10 faster
than \phantomx\ when using 64bit computations and a factor 8.6 faster 
when \phantomx\ is using 32bit computations. The minor difference 
between the 64bit and 32bit versions suggest that the performance difference 
between those two compute modes on the Power8 architecture is minor, especially when 
there is no explicit usage of the vector assembly instructions in the source code.

\begin{figure}
\begin{center}
    \includegraphics[width=0.5\columnwidth]{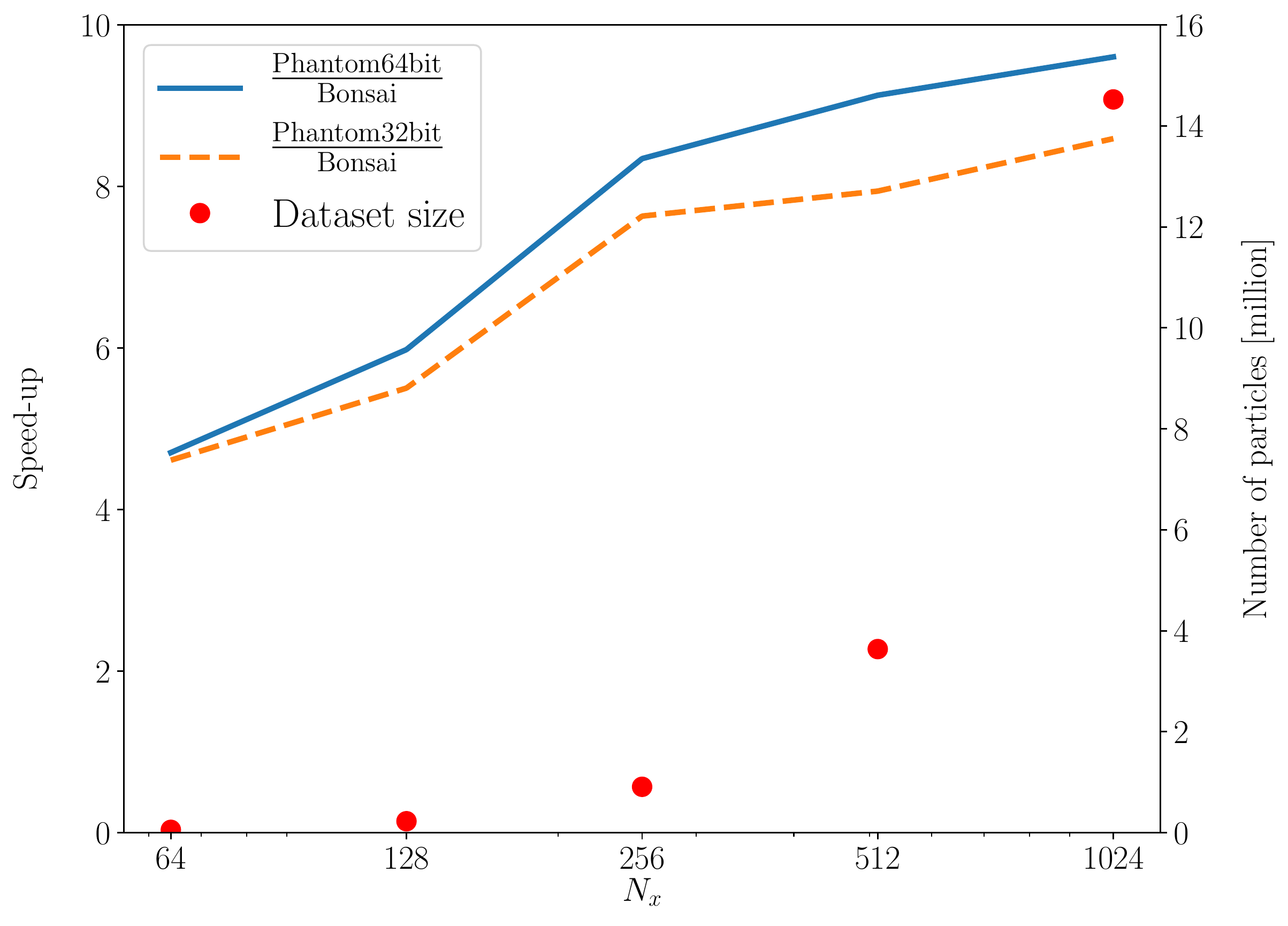}
    \caption{Speed-up when using \bonsai\ vs \phantomx. The $x$-axis
      indicates the Kelvin-Helmholtz dataset size, the left $y$-axis
      indicates the speed-up.  The red circles (right $y$-axis)
      indicate the number of particles in the model. 
      The solid blue (dashed orange) line indicates the difference between \bonsai\ and 
      the 64bit (32bit) version of \phantomx.
      The scaling data is obtained by running the same model using both simulation
      codes for the same amount of simulation time and then comparing
      the average time per iteration step.  }
    \label{fig:KH_performance}
\end{center}    
\end{figure}

\section{Conclusions}
\label{Sect:Conclusion}

In this work we introduced a new GPU accelerated SPH code called
\bonsai.  Our objective was to develop a GPU optimized solver for
fluid dynamics.  We demonstrate that \bonsai\ can compete in terms of
precision and accuracy with state-of-the-art codes when simulating
fluids using the modern SPH equations.  Not only do the results match,
\bonsai\ also executes them up to a factor 10 faster.  This enables
researchers to do more or larger simulations in the same wall-clock
time-frame.  In the same way as we developed the \bonsaig\ pure 
gravitational code, we hope that this allows researchers to perform
simulations using resolutions that where hitherto beyond reach of
modern computers, such as the ones presented in
\cite{2018MNRAS.477.1451F}.

However, we did not develop \bonsai\ as a replacement for stand alone
SPH codes.  Not all the features, such as sub-grid physics, that codes
such as Gadget~\cite{2005MNRAS.364.1105S} and \phantomx\ offer are
implemented. We specifically focused on implementing the fundamental
features that allow faster exploration of parameter space. Once a
final configuration has been found users can opt to run that setting
with a slower, but more versatile code.  Alternatively the user can
combine \bonsai\ with other features via the {\tt AMUSE}
framework~\cite{2009NewA...14..369P, 2013CoPhC.183..456P,
  2013A&A...557A..84P,zwart2018astrophysical}. This allows fast
prototyping while still benefiting from the fast execution of the
density and force computations.

Future work that we plan ourselves are the optimization of the
multi-GPU code path. Currently the code is able to take advantage of
multiple-GPUs, but there is barely and performance improvement because
the CPU part of the code slows the calculation down.  However, if one
wants to run models that do not fit in the memory of a single GPU then
this can be done already with the currently published version.  Other
relatively easy features that could be added are additional time-step
switches, stopping conditions, or a more optimal, per particle group,
determination of the number of density iterations required.

\section*{Acknowledgments}

We thank Terrence Tricco for discussions and Daniel Price and collaborators for making
\phantomx\ open source which helped us during and after the development
of \bonsai\ to verify the correctness of our implementation.  We
further like to thank Evghenii Gaburov, Inti Pelupessy and Natsuki
Hosono for discussions and advice.  This work was supported by the
Netherlands Research School for Astronomy (NOVA), NWO (grant \#
621.016.701 [LGM-II]) and by the European Union's Horizon 2020
research and innovation program under grant agreement No 671564
(COMPAT project).


%
%

\bibliography{bonsai_sph}

\nolinenumbers

\end{document}